\documentstyle[amssymb,12pt,thmsa,sw20aip]{article}

\input{tcilatex}
\begin{document}

\author{Simonetta Frittelli$^{a,b}$ and Ezra T. Newman$^{b}$ \\
$^{a}$Dept$.$ of Physics, \\
Duquesne University, Pgh., PA 15282\\
$^{b}$Dept$.$ of Physics and Astronomy, \\
University of Pittsburgh, Pgh., PA 15260}
\title{An Exact Universal Gravitational Lensing Equation }
\date{Aug.27, 1998 }
\maketitle

\begin{abstract}
We first define what we mean by gravitational lensing equations in a general
space-time. A set of exact relations are then derived that can be used as
the gravitational lens equations in all physical situations. The caveat is
that into these equations there must be inserted a function, a two-parameter
family of solutions to the eikonal equation, not easily obtained, that codes
all the relevant (conformal) space-time information for this lens equation
construction. Knowledge of this two-parameter family of solutions replaces
knowledge of the solutions to the null geodesic equations.

The formalism is then applied to the Schwarzschild lensing problem.
\end{abstract}

\section{Introduction}

The purpose of this note is to first {\it define, }on an arbitrary
Lorentzian space-time{\it ,} what we mean by exact gravitational lensing
equations and then derive a version of these exact equations (which we
believe should be of universal applicability). Our definition reduces to the
more usual lens equations\cite{Ehlers} when we consider approximations;
namely small observation angle and perturbation calculations off either flat
or Robertson-Walker space-times. The basic idea will be to consider the one
parameter family of {\it past} light-cones with apex on a time-like
world-line that should be considered as the history of an observer; i.e., it
defines what an observer can ``see'' and when the observer can see it. The
gravitational lens equations are to be, loosely stated, the space-time
positions of potential light sources, given in terms of the position and
time of the observation and the direction of observation (direction on the
observers past tangent space light-cone).

In Sec.II we will make precise what we mean by the ``lens equation'' and in
Sec.III, using techniques obtained from V.I. Arnold's \cite
{Arnold,Arnold1,Arnold2,Arnold3} theory of Lagrangian and Legendre maps, we
derive a set of general relations, ``lens equations'', that with appropriate
linearization reduce to the standard approximate lens equation\cite{Ehlers}.

A caveat should be expressed here. First of all in any form of lensing
equations one must be looking at the past null geodesics and one thus must 
{\it have the metric tensor} (needed only up to a conformal factor since
only {\it null} geodesics are relevant) in order to calculate, from the
geodesic equations, the {\it null geodesics themselves. }Though in our
version, this information clearly must still be there, its specific
appearance is circumvented by the assumed knowledge of a two-parameter
family of solutions, $u=Z(x^{a},\zeta ,\overline{\zeta }),$ to the eikonal
equation:

\begin{equation}
g^{ab}(x^{a})\partial _{a}Z\partial _{b}Z=0.  \label{1}
\end{equation}
The two parameters, $(\zeta ,\overline{\zeta }),$ are the complex
stereographic coordinates on the sphere. The (null) gradient of $%
Z(x^{a},\zeta ,\overline{\zeta })$ at $x^{a},$ namely $p_{a}=\partial %
_{a}Z(x^{a},\zeta ,\overline{\zeta }),$ sweeps out the sphere of null
directions as $(\zeta ,\overline{\zeta })$ ranges over the sphere. The
function $Z(x^{a},\zeta ,\overline{\zeta }),$ which encodes all the
conformal information of the space-time, is not easily obtained. It can be
found exactly in spaces with sufficient symmetries, e.g., all conformally
flat spaces, (Robertson-Walker), Schwarzschild, Kerr, etc., but in most
applications it will have to be found by perturbation methods. In any case,
our lensing equations will be given explicitly and exactly in terms of $%
Z(x^{a},\zeta ,\overline{\zeta })$ though when applied to physical
situations, approximations will have to be made.

Our starting point will be to assume that we have globally (perhaps given in
different patches) the level surfaces of the function, $u=Z(x^{a},\zeta ,%
\overline{\zeta }).$

\begin{remark}
For simplicity of discussion, we have assumed that the solutions to the
eikonal equation will have been given by a function of the local
coordinates, $x^{a}.$ Actually there might be singular regions (caustics) or
self-intersections, where this is not possible; the surfaces could then be
given in parametric form. We will nevertheless use the language of the
simple function, $Z(x^{a},\zeta ,\overline{\zeta }).$
\end{remark}

\section{Defining the Lens Equation}

We first assume that our space-time allows the null geodesics to be
infinitely extendable into both the future and the past. In any region local
coordinates will exist such that the world-line, ${\frak L,}$ of an observer
can be given parametrically by

\begin{equation}
x^{a}=X_{0}^{a}(\tau )  \label{2}
\end{equation}
and the one parameter family of light-cones,${\frak C}(X_{0}^{a}(\tau )%
{\frak ),}$ by

\begin{equation}
x^{a}=X^{a}(X_{0}^{a}(\tau ),\eta ,\bar{\eta},s)\equiv X^{a}(\tau ,\eta ,%
\bar{\eta},s)  \label{3}
\end{equation}
where $(\eta ,\bar{\eta})$ label the sphere's worth of null directions at $%
x^{a}=X_{0}^{a}(\tau )$ and $s$ is an affine parameter along each of the
geodesics labeled by $(\tau ,\eta ,\bar{\eta}).$ $L^{a}=\partial _{s}X^{a}$ $%
\neq 0$ is the (null) tangent vector to the geodesics. \{Derivatives of $%
X^{a}$ with respect to the other parameters yield connecting vectors. They
will not directly interest us here though they often are used in the study
of the caustics of the light-cones, ${\frak C}(X_{0}^{a}(\tau ){\frak ).}$
We will approach the caustic issue from a slightly different point of
view.\} We further assume that the local coordinates $x^{a}$ are such that
three of them, $x^{i},$ are space-like and one, $x^{0},$ is time-like. From
this and from the non-vanishing null character of $L^{a},$ using the
implicit function theorem, one (at least one) of the space-like equations (%
\ref{3}) can be solved for $s.$ Let us call that special coordinate $x^{*}$
and the other two space-like coordinates, $x^{\alpha }.$ We then have that

\begin{equation}
s=S(x^{*},\tau ,\eta ,\bar{\eta})  \label{4}
\end{equation}
and the remaining three Eqs.(\ref{3}) become, when $s$ is replaced by $%
S(x^{*},\tau ,\eta ,\bar{\eta})$

\begin{eqnarray}
x^{0} &=&X^{0}(\tau ,\eta ,\bar{\eta},S(x^{*},\tau ,\eta ,\bar{\eta})\equiv 
\widehat{X}\!^{0}(x^{*},\tau ,\eta ,\bar{\eta}),  \label{5} \\
x^{\alpha } &=&X^{\alpha }(\tau ,\eta ,\bar{\eta},S(x^{*},\tau ,\eta ,\bar{%
\eta}))\equiv \widehat{X}\!^{\alpha }(x^{*},\tau ,\eta ,\bar{\eta}).
\label{6}
\end{eqnarray}

If we think of a light source at the spatial position ($x^{*},x^{\alpha })$
emitting light at time $x^{0},$ then Eq.(\ref{5}) relates the time of
emission $x^{0}$ at $x^{*},$ with the time of arrival $\tau $ and arrival
direction, $(\eta ,\bar{\eta}).$ We will refer to Eq.(\ref{5}) as the time
of arrival relation or function.\qquad \qquad \qquad

The {\it lens equation} is {\it defined }by Eq.(\ref{6}){\it ; } it
expresses two of the spatial coordinates, $x^{\alpha },$ in terms of the
arrival time at the observer, $\tau ,$ the direction at the observer, that
the observer sees the source and {\it one} of the source coordinates. Most
often in applications the two coordinates $x^{\alpha }$ are identified with
some angular position coordinates ($\theta ,\varphi $) of the source and the 
$x^{*}$ as some form of ``radial'' or ``distance from observer coordinate'',
say $D$. In this case the lens equation reads

\begin{eqnarray}
\theta &=&\Theta (D,\tau ,\eta ,\bar{\eta}),  \label{7} \\
\varphi &=&\Phi (D,\tau ,\eta ,\bar{\eta}).  \nonumber
\end{eqnarray}
Much of conventional lensing theory is based on finding approximate versions
of Eq.(\ref{7}). An important issue is the inversion of these equations,
i.e., finding the observation angles in terms of the source position, namely
when can one rewrite them as

\begin{equation}
(\eta ,\bar{\eta})=(N(D,\tau ,\theta ,\varphi ),\text{ }\overline{N}(D,\tau
,\theta ,\varphi ))  \label{8}
\end{equation}
and the multiplicity of these solutions. The condition for caustics on the
past light-cone at $D$ (the non-invertibility of (\ref{7})) is the vanishing
of the Jacobian of Eq.(\ref{7});

\begin{equation}
J=\frac{\partial (\theta ,\varphi )}{\partial (\eta ,\bar{\eta})}.  \label{9}
\end{equation}
We return to this issue in the next Section in a more general context.

\section{The Construction of the Past Light-Cone}

The construction of the light-cones, ${\frak C}(X_{0}^{a}(\tau ){\frak ),}$
Eq.(\ref{3}) and hence, the time of arrival function and the lens equations,
Eqs.(\ref{5}) and (\ref{6}) was based on the integration of the null
geodesic equations. In this section we will give an alternative construction
of Eqs.(\ref{5}) and (\ref{6}) based on the solution to the eikonal equation.

As discussed in the introduction, we assume that we have the two-parameter
family of solutions, $u=Z(x^{a},\zeta ,\overline{\zeta }),$ to the eikonal
equation where for each value of $(\zeta ,\overline{\zeta })$ the level
surfaces of $Z$ are null surfaces, i.e., $\partial _{a}Z$ is a null
covector. Furthermore it is assumed that at each point $x^{c},\partial _{a}Z$
sweeps out the entire null cone at $x^{a}$ as $(\zeta ,\overline{\zeta })$
goes thru its range, $S^{2}$.

\begin{remark}
We repeat and emphasize that the level surfaces of the solutions to Eq.(\ref
{1}) though referred to as ``null or characteristic surfaces'' are not
strictly speaking surfaces; they can have self-intersections and in general
are only piece-wise smooth. Though Arnold\cite{Arnold} refers to them as
``big-wave-fronts'' we will continue to call them null surfaces. The
intersection of a big wave front with a generic three surface yields a
two-dimensional (small) wave front.
\end{remark}

The first thing that we want to show is that the light-cones, ${\frak C}
(X_{0}^{a}(\tau ){\frak )},$ with apex on an arbitrary curve, ${\frak L,}$
can be constructed from knowledge of the function $Z$.\qquad

One sees immediately, from the eikonal equation, that the function 
\begin{equation}
F(x^{a},\tau ,\zeta ,\overline{\zeta })\equiv Z(x^{a},\zeta ,\overline{\zeta 
})-Z(X_{0}^{a}(\tau ),\zeta ,\overline{\zeta })=0,  \label{c}
\end{equation}
for {\it each fixed value} of $\tau ,$defines a two-parameter family of
surfaces which all pass thru the point $X_{0}^{a}(\tau )$ and which are all
null surfaces.

\begin{remark}
Note that $\partial _{\tau }F\equiv -\partial _{\tau }Z=-V^{a}(\tau
)\partial _{a}Z\neq 0,$ with $V^{a}(\tau )=\partial _{\tau }X_{0}^{a}(\tau
). $ The non-vanishing of $\partial _{\tau }F$ is because the scalar product
of a non-vanishing null vector with a non-vanishing time-like vector is
different from zero. It now follows from the implicit function theorem that
Eq.(\ref{c}) is the implicit version of the function $\tau =T(x^{a},\zeta ,%
\overline{\zeta }),$ i.e., we have 
\begin{equation}
Z(x^{a},\zeta ,\overline{\zeta })-Z(X_{0}^{a}(\tau ),\zeta ,\overline{\zeta }%
)=0\Leftrightarrow \tau =T(x^{a},\zeta ,\overline{\zeta }).  \label{c*}
\end{equation}
The function $T(x^{a},\zeta ,\overline{\zeta })$ (or $F)$ defines, what
Arnold\cite{Arnold1,Arnold3} calls a generating family and is used to
construct, via its envelope, the light-cone for each value of $\tau .$
\end{remark}

Specifically the envelope of this family is constructed by demanding that%
\cite{landau} 
\begin{eqnarray}
\partial _{\zeta }F(x^{a},\tau ,\zeta ,\overline{\zeta }) &=&0,  \label{d} \\
\partial _{\overline{\zeta }}F(x^{a},\tau ,\zeta ,\overline{\zeta }) &=&0 
\nonumber
\end{eqnarray}
where ($\partial _{\zeta },\partial _{\overline{\zeta }}$ ) denote the
derivatives with respect to the ($\zeta ,\overline{\zeta }).$ Assuming for
the moment that (\ref{d}) could be solved for the ($\zeta ,\overline{\zeta }%
)=(\Gamma (x^{a},\tau ),\overline{\Gamma }(x^{a},\tau ))$ then when they are
substituted into (\ref{c} ) one obtains the function 
\begin{equation}
\hat{F}(x^{a},\tau )\equiv F(x^{a},\tau ,\Gamma (x^{a},\tau ),\overline{%
\Gamma }(x^{a},\tau ))=Z(x^{a},\Gamma ,\overline{\Gamma })-Z(X_{0}^{a}(\tau
),\Gamma ,\overline{\Gamma })=0.  \label{e}
\end{equation}

Once again using the implicit function theorem with Eq.(\ref{d}), Eq.(\ref{e}
), can be written as

\begin{equation}
\hat{F}(x^{a},\tau )\equiv Z(x^{a},\Gamma ,\overline{\Gamma }
)-Z(X_{0}^{a}(\tau ),\Gamma ,\overline{\Gamma })=0\Leftrightarrow \tau =\hat{%
T}(x^{a})  \label{e*}
\end{equation}

Using either $\hat{F}(x^{a},\tau )=0$ or $\tau =$ $\hat{T}(x^{a})$ one can
easily see that they define a one parameter, ($\tau ),$ family of null
surfaces (and satisfy the eikonal equation); this follows from either $%
\partial _{a}\hat{F}(x^{a},\tau )|_{\tau }=\partial _{a}Z(x^{a},\zeta ,%
\overline{\zeta })$ or from $\partial _{a}T(x^{a})\varpropto \partial
_{a}Z(x^{a},\zeta ,\overline{\zeta }).$ Furthermore, at $x^{a}=X_{0}^{a}(%
\tau ),$ the gradients $p_{a}=\partial _{a}Z(X_{0}^{a}(\tau ),\zeta ,%
\overline{\zeta })$ sweep out all null directions. $[$At $%
x^{a}=X_{0}^{a}(\tau ),$ Eq.(\ref{d}) is identically satisfied and can not
be solved for the ($\zeta ,\overline{\zeta })=(\Gamma (x^{a},\tau ),%
\overline{\Gamma }(x^{a},\tau ));$ all values of ($\zeta ,\overline{\zeta })$
are allowed. See discussion below.] We thus see that $\hat{F}(x^{a},\tau )=0$
[or $\tau =\hat{T}(x^{a})]$ represents the family of null cones, ${\frak C}
(X_{0}^{a}(\tau )),$ with apex along ${\frak L.}$

The assumption that Eq.(\ref{d}) could be solved for ($\zeta ,\overline{
\zeta })=(\Gamma (x^{a},\tau ),\overline{\Gamma }(x^{a},\tau ))$ depended on
the non-vanishing of the determinant $J$ of the matrix

\begin{equation}
J_{ij}\equiv \left\| 
\begin{array}{cc}
\partial _{\zeta }\partial _{\zeta }F & \partial _{\overline{\zeta }
}\partial _{\zeta }F \\ 
\partial _{\overline{\zeta }}\partial _{\zeta }F & \partial _{\overline{%
\zeta }}\partial _{\overline{\zeta }}F
\end{array}
\right\|  \label{e**}
\end{equation}
$J$ does vanish at the singularities of the ``surface'' $F(x^{a},\tau ,\zeta
,\overline{\zeta }),$ e.g., at the apex $x^{a}=x_{0}^{a},$ among other
regions. In fact the vanishing of $J$ {\it defines\cite{Arnold1,Arnold2}}
the caustics of the family of the cones, ${\frak C}(X_{0}^{a}(\tau )).$

In general, whether or not $J=0$, Eqs.(\ref{d}) and (\ref{c}) can be solved
for other variables, namely {\it some }set of{\it \ }three (say $x^{I},$
which might be different in different regions) of the four $x^{a},$ in terms
of the fourth one (say $x^{\#}),$ $\tau $ and the ($\zeta ,\overline{\zeta }%
),$ i.e., 
\begin{equation}
x^{I}=x^{I}(x^{\#},\tau ,\zeta ,\overline{\zeta }).  \label{f}
\end{equation}
Note the important point that if the coordinates $x^{a}$ are such that three
of them (say $x^{j}=\{x^{\alpha },x^{*}\})$ are space-like and one of them
is a time coordinate, $x^{0},$ then Eq.(\ref{f}) has a stronger version,
namely

\begin{eqnarray}
x^{0} &=&\widehat{X}\!^{0}(x^{*},\tau ,\zeta ,\overline{\zeta }),  \label{f*}
\\
x^{\alpha } &=&\widehat{X}\!^{\alpha }(x^{*},\tau ,\zeta ,\overline{\zeta })
\label{f**}
\end{eqnarray}
where the two $x^{j}$ and the $x^{*}$ are the three space-like coordinates.
That one can solve for the $x^{0}=\hat{X}^{0}(\tau ,x^{*},\zeta ,\overline{%
\zeta })$ follows from the implicit function theorem and from the fact that $%
\hat{T}(x^{a})$ satisfies the eikonal equation and hence $\partial \hat{T}%
(x^{a})/\partial x^{0}\neq 0.$

Eqs.(\ref{f*}) and (\ref{f**}) are a parametric representation of ${\frak C}
(X_{0}^{a}(\tau ))$ via the null geodesics that rule it. For the different
given values of the ($\zeta ,\overline{\zeta }),$ they are the null
geodesics thru $x_{0}^{a}.$ They are the {\it explicit version} of the
implicit relations of Eqs.(\ref{c}) and (\ref{d}).

We thus see that Eqs.(\ref{c}) and (\ref{d}) are equivalent to the lens and
the time of arrival equations, (\ref{5}) and (\ref{6}) of the previous
section.

\section{Examples; Flat-Space and Schwarzschild Space-Time}

First, as a rather trivial example, using a family of $Z(x^{a},\zeta ,%
\overline{\zeta })$ in flat space, we construct the family ${\frak C}%
(X_{0}^{a}(\tau )).$ A particularly useful two-parameter family of solutions
to the flat-space eikonal equation are the family of all plane-waves with
can be represented\cite{STG} in the following fashion;

\begin{equation}
Z(x^{a},\zeta ,\overline{\zeta })=x^{a}\ell _{a}(\zeta ,\overline{\zeta })
\label{natural}
\end{equation}
where $\partial _{a}Z=\ell _{a}(\zeta ,\bar{\zeta})$ represents the
covariant version of the {\it null} vectors $\ell ^{a},$ pointing in all
possible directions, with Cartesian components given as 
\begin{equation}
\ell ^{a}(\zeta ,\overline{\zeta })=\frac{1}{\sqrt{2}(1+\zeta \overline{%
\zeta })}\left( (1+\zeta \overline{\zeta }),(\zeta +\overline{\zeta }),i(%
\overline{\zeta }-\zeta ),(\zeta \overline{\zeta }-1)\right) .  \label{A}
\end{equation}
Using the null basis set, (for each value of $(\zeta ,\overline{\zeta })),$
\{$\ell ^{a}(\zeta ,\overline{\zeta }),$ $m^{a}(\zeta ,\overline{\zeta }),$ $%
\overline{m}^{a}(\zeta ,\overline{\zeta }),$ $n^{a}(\zeta ,\overline{\zeta }%
)\}$ where

\begin{eqnarray}
m^{a} &=&(1+\zeta \overline{\zeta })\partial _{\zeta }\ell ^{a},\qquad 
\overline{m}^{a}=(1+\zeta \overline{\zeta })\partial _{\overline{\zeta }%
}\ell ^{a},  \label{B} \\
n^{a} &=&\ell ^{a}+(1+\zeta \overline{\zeta })^{2}\partial _{\zeta }\partial
_{\overline{\zeta }}\ell ^{a}  \nonumber
\end{eqnarray}

that have all vanishing scalar products except for $\ell ^{a}n_{a}=-m^{a}%
\bar{m}_{a}=1.$

The lens and time of arrival equations, i.e., Eqs.(\ref{c}) and (\ref{d}),
are then

\begin{eqnarray}
\qquad (x^{a}-X^{a}(\tau ))\ell _{a}(\zeta ,\overline{\zeta }) &=&0,
\label{C} \\
(x^{a}-X^{a}(\tau ))m^{a} &=&0,\qquad (x^{a}-X^{a}(\tau ))\overline{m}^{a}=0
\label{D}
\end{eqnarray}
where $x^{a}=X^{a}(\tau ),$ is an observers world-line, ${\frak L.}$ Eqs.(%
\ref{C}) and (\ref{D}), using (\ref{B}), are easily solved for a parametric
form of the light-cone as

\begin{equation}
x^{a}=X^{a}(\tau )+s\ell ^{a}(\zeta ,\overline{\zeta }),\text{ }  \label{E}
\end{equation}
with $s$ a parameter along the null geodesic.. If the observation is to be
at an angle close to (say) the $x^{*}=x^{3}-axis,$ then $s=(x^{3}-X^{3}(\tau
))/\ell ^{3}(\zeta ,\bar{\zeta})$ and eliminating $s$ from the remaining
equations (\ref{E}), we have for (\ref{f*}) and (\ref{f**});

\begin{equation}
x^{0}=X^{0}(\tau )+(x^{3}-X^{3}(\tau ))\frac{\ell ^{0}(\zeta ,\overline{%
\zeta })}{\ell ^{3}(\zeta ,\overline{\zeta })},  \label{F*}
\end{equation}

\begin{eqnarray}
x^{1} &=&X^{1}(\tau )+(x^{3}-X^{3}(\tau ))\frac{\ell ^{1}(\zeta ,\overline{%
\zeta })}{\ell ^{3}(\zeta ,\overline{\zeta })},  \label{F**} \\
x^{2} &=&X^{2}(\tau )+(x^{3}-X^{3}(\tau ))\frac{\ell ^{2}(\zeta ,\overline{%
\zeta })}{\ell ^{3}(\zeta ,\overline{\zeta })}  \nonumber
\end{eqnarray}
the trivial time of arrival function and lens equation. A simple calculation
of the Jacobian,

\begin{equation}
J=\partial (x^{1},x^{2})/\partial (\zeta ,\overline{\zeta })
\end{equation}
verifies that it is different from zero and hence in the flat case there are
no caustics.

There are two ways to find the equations equivalent to Eqs.(\ref{F*}) and (%
\ref{F**}) in a Schwarzschild space-time; one could integrate the eikonal
equation directly, (this is easily doable by separation of variables\cite
{landau2}) or by integration of the null geodesic equations with appropriate
initial conditions. (Actually this integration is most easily performed by
the use of Hamilton-Jacobi techniques via the Eikonal equation.) In a recent
work by Tom Kling, the exact Schwarzschild lensing equations were found\cite
{kling}. We simply quote these results.

Due to the spherical symmetry of the Schwarzschild lensing problem, with no
loss of generality, one could place the observer at an arbitrary point on a
fixed radial line - we chose the negative $z$ axis - and place the sources
in a special plane - the $x-z$ plane. In polar coordinates that means the
observer is at the angular position $\theta _{0}=\pi ,$ $\phi _{0}=0.$ (For
simplicity of discussion, since we working in a fixed plane, we let $0\leq
\theta \leq 2\pi ,$ $\phi =0,$ rather than $0\leq \theta \leq \pi $ with $%
\phi =0$ and $\pi .).$ The angular position of a source is then given by $%
0\leq \theta \leq 2\pi $ and $\phi =0.$ Rather then the usual Schwarzschild
radial coordinate $r,$ it is much more useful to use

\begin{equation}
l~=\frac{1}{\sqrt{2}r}~  \label{G}
\end{equation}
and the retarded time coordinate, $u=t-r-2Mlog(r-2M)$

The time of arrival equation and the lens equation have the form

\begin{equation}
u=u_{0}-U(l_{0},\psi ,l)  \label{H}
\end{equation}

\begin{equation}
\theta =\Theta (l_{0},\psi ,l)  \label{H*}
\end{equation}
where ($l_{0},u_{0})$ are the position coordinates of the observer, who is
assumed to be at the fixed position $l_{0}$ on the negative $z$- axis. The
coordinates of the source are ($\theta ,u,l)$ and $\psi $ is the observation
angle, namely the angle between the incoming null geodesic projected into
the rest-frame of the observer, and the inward directed radial vector at the
observer, i.e., in the positive $z$ direction. The roles of $\tau $ and x$%
^{*},$ in Eqs.(\ref{f*}) and (\ref{f**}), are played here by $u_{0}$ and $l.$
The exact expressions for $U$ and $\Theta $ are given by the integral
expressions,

\begin{equation}
u=u_{0}-\int_{l_{0}}^{\hat{l}}\frac{dl^{\prime }(1{+}\sqrt{1-\frac{\sin
^{2}\psi ~l^{\prime }{}^{2}~f(l^{\prime })}{l_{0}^{2}~f(l_{0})})}}{
2l^{^{\prime }2}f(l^{^{\prime }})\sqrt{1-\frac{\sin ^{2}\psi ~l^{\prime
}{}^{2}~f(l^{\prime })}{l_{0}^{2}~f(l_{0})}}}+\int_{\hat{l}}^{l}\frac{
dl^{\prime }(1{-}\sqrt{1-\frac{\sin ^{2}\psi ~l^{\prime }{}^{2}~f(l^{\prime
})}{l_{0}^{2}~f(l_{0})})}}{2l^{^{\prime }2}f(l^{^{\prime }})\sqrt{1-\frac{%
\sin ^{2}\psi ~l^{\prime }{}^{2}~f(l^{\prime })}{l_{0}^{2}~f(l_{0})}}}
\label{I}
\end{equation}

\begin{equation}
\Theta (l_{0},\psi ,l)=\pi -\frac{\sin \psi }{l_{0}~\sqrt{f(l_{0})}}
~\int_{l_{0}}^{\hat{l}}~\frac{dl^{\prime }}{\sqrt{(1-\frac{\sin ^{2}\psi
~l^{\prime }{}^{2}f(l^{\prime })}{l_{0}^{2}~f(l_{0})})}}+\frac{\sin \psi }{%
l_{0}~\sqrt{f(l_{0})}}~\int_{\hat{l}}^{l}~\frac{dl^{\prime }}{\sqrt{(1-\frac{%
\sin ^{2}\psi ~l^{\prime }{}^{2}f(l^{\prime })}{l_{0}^{2}~f(l_{0})})}}
\label{I*}
\end{equation}
where

\begin{equation}
f(l)=1-\frac{2M}{r}=(1-2\sqrt{2}Ml)  \label{J}
\end{equation}
and $\hat{l}$ is the positive root of the cubic equation

\begin{equation}
{(\sin \psi )^{2}}\hat{l}^{2}(1-2\sqrt{2}M\hat{l})-l_{0}^{2}~(1-2\sqrt{2}%
Ml_{0})=0.  \label{K}
\end{equation}

\qquad If the source is situated on the positive $z$- axis, i.e., at $\theta
=0$ then Eq.(\ref{I*}) becomes

\begin{equation}
0=\pi -\frac{\sin \psi }{l_{0}~\sqrt{f(l_{0})}}~\int_{l_{0}}^{\hat{l}}~\frac{%
dl^{\prime }}{\sqrt{(1-\frac{\sin ^{2}\psi ~l^{\prime }{}^{2}f(l^{\prime })}{%
l_{0}^{2}~f(l_{0})})}}+\frac{\sin \psi }{l_{0}~\sqrt{f(l_{0})}}~\int_{\hat{l}%
}^{l}~\frac{dl^{\prime }}{\sqrt{(1-\frac{\sin ^{2}\psi ~l^{\prime
}{}^{2}f(l^{\prime })}{l_{0}^{2}~f(l_{0})})}}  \label{L}
\end{equation}
which can be considered as an implicit function, $F(\psi ,l,l_{0}))=0,$ that
defines the observation angle $\psi $ when the source and observer are
co-linear with the origin; $\psi $ is then the observation angle of the
Einstein ring.

Note that in this case we have cheated slightly in that we have used null
coordinates for the time of arrival and lens equations rather than the
spatial/time coordinates of Secs. II and III.

Simple approximations\cite{kling} to Eqs.(\ref{I}), (\ref{I*}) and (\ref{L})
yield the standard linearized Schwarzschild time of arrival/lens equation
and Einstein angle.

\section{Discussion}

In Sec.II, we gave a general definition of an exact time of arrival/lens
equation that, in principle, is applicable to all physical situations. Its
application depended on knowing how to construct the past light-cones from
an observers world-line or on the construction of a two-parameter family of
solutions to the eikonal equation.

We are studying the possibility of applying these techniques to statistical
perturbations of the homogeneous, isotropic cosmological models for which
the eikonal solutions are known.

From the point of view of contact with observation, it appears extremely
unlikely that the exact equations discussed here, and their associated
corrections from the linearized approach to lensing, implicit in the exact
equations, will have observational consequences in the near future.
Nevertheless, the extraordinarily rapid advances in observational techniques
must make one dubious of absolute statements that something is unobservable.
To emphasize this point we end with a quote from Einstein's 1936 paper\cite
{AE} on lensing;

\begin{itemize}
\item  \qquad {\it ``....there is no great chance of observing this
phenomenon.''}
\end{itemize}

\section{Acknowledgments}

The authors thank the NSF for financial support under grant \#PHY-9722049.
We also extend our appreciation to Tom Kling for generously allowing us to
summarize his results on Schwarzschild lensing


\section{Bibliography}

\begin{thebibliography}{99}
\bibitem{Ehlers}  P. Schneider, J. Ehlers, E.E. Falco, {\it Gravitational
Lenses}, Springer-Verlag, New YOrk, Berlin, Heidelberg, 1992

\bibitem{Arnold}  {V.I Arnold, {\it Catastrophe Theory}, Springer Verlag,
Berlin, Heidelberg, NY, Tokyo, (1986). }

\bibitem{Arnold1}  V. I. Arnold, S.P. Novikov, eds., {\it Dynamical Systems; 
} Vol. IV, , Springer-Verlag, Berlin, Heidelberg, NY, (1990).

\bibitem{Arnold2}  {V.I Arnold, {\it Mathematical Methods of Classical
Mechanics}, Springer Verlag, Berlin, Heidelberg, NY, Tokyo, (1980). }

\bibitem{Arnold3}  V. I. Arnold, S.M. Gusein-Zade, A. N.. Varchenko, {\it %
Singularities of Differentiable Maps}, Vol. I, Birkhauser, Boston, Basel,
Stuttgart, 1985.

\bibitem{landau}  {L. Landau and Lifschitz, {\it Classical Mechanics},
Pergamon Press, Headington Hill Hall, Oxford, 4, 5 Fitzroy Sq. London, W.1,
(1960).}

\bibitem{STG}  {Simonetta Frittelli, Ezra T. Newman and Gilberto
Silva-Ortigoza, }{\it The Eikonal Equation in Flat Space; Null Surfaces and
Their Singularities}{, accepted for publication in JMP.}

\bibitem{landau2}  {L. Landau and Lifschitz, {\it Classical Theory of
Fields, }Pergamon Press, Headington Hill Hall, Oxford, 4, 5 Fitzroy Sq.
London, W.1, (1962).}

\bibitem{kling}  Thomas P. Kling, Ezra T. Newman, {\it Null Surfaces in
Spherically Symmetric Space-Times, }submitted Phys. Rev.D

\bibitem{AE}  Einstein, A. {\it Lens-Like Action of a Star by the Deviation
of Light in the Gravitational Field, }Science, Vol 84, p506, (1936)
\end{thebibliography}
\end{document}